\documentclass{aastex}

\newcommand\msun{$M_{\sun}$}

\newcommand\ha{H${\alpha}$~}

\slugcomment{}
\shortauthors{Stanghellini et al.}
\shorttitle{Large Magellanic Cloud Planetary Nebulae} 

\begin{document}

\title{Large Magellanic Cloud Planetary Nebula Morphology:
Probing Stellar Populations and Evolution.
\footnote{Based on observations made with the NASA/ESA Hubble Space Telescope,
and from the {\it HST} Data Archive, 
obtained at the Space Telescope Science Institute, which is operated 
by the Association of universities for research in Astronomy, Inc., under
NASA contract NAS 5--26555}}

\author{Letizia Stanghellini\altaffilmark{2,3}, Richard A. Shaw}
\affil{Space Telescope Science Institute, 3700 San Martin Drive,
Baltimore, Maryland 21218, USA}

\author{Bruce Balick}
\affil{Department of Astronomy, University of Washington, Seattle, 
Washington 98195}
\and
\author{J. Chris Blades}
\affil{Space Telescope Science Institute}

\altaffiltext{2}{Affiliated to the Astrophysics Division, Space Science
Department of ESA}
\altaffiltext{3}{{\it on leave,} Osservatorio Astronomico di Bologna}

\begin{abstract} 
Planetary Nebulae (PNe) in the Large Magellanic Cloud (LMC) offer the unique 
opportunity to study both the Population and evolution of low- and intermediate-mass 
stars, by means of the morphological type of the nebula. Using 
observations from our LMC PN morphological survey, and including 
images available in the {\it HST} Data Archive, and published chemical abundances, 
we find that asymmetry in PNe is strongly correlated with a younger 
stellar Population, as indicated by the abundance of elements that are 
unaltered by stellar evolution (Ne, Ar, S). While similar results have 
been obtained for Galactic PNe, this is the first demonstration of the 
relationship for extra-galactic PNe. 
We also examine the relation between morphology and abundance of the products 
of stellar evolution. We found that 
asymmetric PNe have higher nitrogen and lower carbon abundances than 
symmetric PNe. Our two main results are broadly consistent with the 
predictions of stellar evolution if the progenitors of asymmetric PNe 
have on average larger masses than the progenitors of symmetric PNe. The 
results bear on the question of formation mechanisms for asymmetric PNe, 
specifically, that the genesis of PNe structure should relate strongly 
to the Population type, and by inference the mass, of the progenitor star, 
and less strongly on whether the central star is a member of a close binary system. 

\end{abstract} 

\keywords{Stars: AGB and post-AGB --- stars: evolution --- planetary nebulae:
general --- Magellanic Clouds}

\section {Introduction} 

Planetary Nebulae (PNe) are produced as stars age. They are 
fundamental to understanding the evolution of 
stars whose initial mass lies below the Supernova limit. Classically, 
Galactic PNe have been divided in Population classes according to their 
spatial distribution, kinematics, and chemical content 
\citep{gre72,pei78,mac99}. It appears that there are clearly different PN 
Populations in the Galaxy, from old disk Population PNe (Peimbert's Type I) 
to extreme Pop II PNe, located in the 
Galactic halo or in the bulge (Peimbert's Type IV and V). 
The PN morphology varies systematically across these classes \citep{pei97}, 
and most of the Type I PNe are asymmetric in shape. 

More recent studies 
of large Galactic PNe samples have shown that, to first approximation, the 
morphology of PNe is linked to the spatial distribution within the Galaxy, 
and to the mass of the progenitor star \citep{sta93,man00}. 
The fact that most bipolar and quadrupolar PNe lie on average closer to 
the Galactic plane than round and elliptical PNe, and that highly asymmetric 
PNe appear to host the most massive central stars (CSs), suggests that 
asymmetric PNe are the likely progeny of a younger stellar Population than 
symmetric PNe.\footnote
{In this paper we group elliptical and round PNe in the {\it symmetric} 
class, and bipolar and bipolar core PNe in the {\it asymmetric} class.
Quadrupolar PNe in the LMC may have questionable morphology, thus we 
segregate them from these two major classes. Note that this definition of 
symmetric and asymmetric PNe is identical to the {\it elliptical} and 
{\it bipolar} classifications in both Stanghellini et al. (1993) 
and \citet{cor95}.} 

An important and long-standing astrophysical issue is the degree to which PNe enrich heavy elements in the ISM. In the Galaxy, PNe supply almost an
order of magnitude more mass per year than supernovae \citep{ost89}. 
Depending on the progenitor's mass, PNe are expected to enrich the ISM
with carbon and nitrogen \citep{ibe83}. PNe stellar progenitors
undergo several dredge ups (e.g., Iben \& Renzini 1983), some of which
enrich the surfaces with C and N. Subsequent winds will carry the
enriched gas into the ISM. PNe are known to account for half of the carbon
and most of the nitrogen enrichment in the solar environment 
\citep{hen98}. Therefore a study of C and N abundances relative to
the elements which are not altered during the evolution of PNe
progenitors, such as Ne,
Ar, and S, provides a means for gauging the efficacy of C-N enrichment rates by
PNe.

Abundance studies as a function of Population class and morphological type
have been carried out for Galactic PNe. However, since asymmetric PNe lie close
to the plane where foreground extinction is severe, and since such nebulae are
also formed from the most massive progenitors, 
Galactic PNe suffer a serious selection bias for
understanding enrichment rates by PNe of different types.
Accordingly, we have begun a comprehensive survey of Magellanic Cloud PNe in 
order to investigate these and other parameters in a large and well-understood
sample. We are using {\it HST} and STIS in direct imaging and slitless spectroscopy
mode for this survey, and our first observational results are described in 
\citet{sha00}.
For our purposes the advantage of
LMC PNe is their relatively low foreground extinction (i.e., their 
low selection
bias) as well as their well determined distances. That is, abundance studies
tend to be complete to a limiting nebular luminosity, which is far less of a
bias than a brightness limited survey in the presence of highly variable
extinction with Galactic latitude.

In the study presented here, we have selected
a subsample of LMC PNe based solely on the availability
of morphological information (from {\it HST} images) and relative chemical abundances
(from ground-based spectroscopy in the literature). The {\it HST} data come from 
\citet{sha00}, from the {\it HST} Data Archive, and from the
literature.
 In $\S$2 we
describe the database and the data analysis, and discuss the results; and in $\S$3 we discuss the
uncertainties and possible systematic effects.

\section {PN Morphology, Stellar Populations, and Stellar Evolution}

The morphological data used in this paper consist of all LMC PNe 
observed in imaging mode with {\it HST}. The sample includes the pre-refurbishment 
data set of PC1 and FOC images (see Stanghellini et al.~1999), the unpublished WFPC2 images 
in the Hubble Data Archive ({\it HST} Program ID: 6407, PI: Dopita, Cycle 6),
and the set of 27 LMC PNe observed to date within our {\it HST}/STIS snapshot 
survey \citep{sha00}. 

The abundance data set was compiled from papers by 
\citet{lei96}, \citet{mon88}, \citet{dop97}, and \citet{dop91a,dop91b}. 
We have used  
abundances of carbon, nitrogen, oxygen, argon, neon, and sulfur 
where available. However, some of the quoted abundances are model dependent 
\citep{dop91a,dop91b}; abundances therein were used only 
when the model independent 
values were not available. A discussion of the effects of the data set 
inhomogeneity is presented in the next section, along with a critical  
analysis of the published abundances, including uncertainties 
and systematic effects that 
could affect our conclusions. 

We classified PN morphologies homogeneously according to their shape in 
narrow-band [\ion{O}{3}] $\lambda$5007 images: Round, elliptical, 
bipolar, bipolar core \citep{sta99}, quadrupolar, and point-symmetric 
PNe are included in the sample. 
While classifying the images, we realized that in a few cases the morphologies 
are uncertain. 
These objects are not 
included in the numerical analysis and discussion, nor in the figures. 

Figures 1 and 2 show the distribution of the morphological types with 
respect to neon, argon, and sulphur abundances, which have presumably remain 
unchanged since the birth of the progenitor star. As such, these 
elements should be good indicators of the Population type of the progenitor 
stars. The crossed large circles in these figures marks the location of 
the {\it average} abundance for LMC \ion{H}{2} regions \citep{lei96}.
These and other young population LMC abundances are given in Table
1, discussed below.

The segregation of morphological type with respect to S and to Ne (Fig. 1)
is striking; Ar does not appear to be as good a
discriminant (Fig. 2). Specifically, 
the overwhelming majority of PNe with log Ne + $12 > 7.6$ are asymmetric, 
and none is round. Furthermore, all PNe with log Ne + $12 < 7.6$ are 
symmetric, with the exception of one quadrupolar PN. Evidently, PN 
morphology is a good indicator of progenitor Population in the LMC. 
By comparing 
the Ne abundance in PNe with the average for \ion{H}{2} regions (which gives a
good indication of the abundance of the young stellar Population), 
it is evident that most of the asymmetric PNe are enriched with respect to
the young Population, while the opposite holds for symmetric PNe. 
Sulfur is related to nebular shape in a 
similar way. In fact, only one of the 13 symmetric PNe has S larger 
than the LMC \ion{H}{2} region average, and three asymmetric PNe are 
under-abundant with respect to the \ion{H}{2} regions. 

To quantify the importance of the neon, sulphur, and argon
overabundance in asymmetric PNe, in Table 1 we list the average abundances 
for our PN sample, together with the abundances of other significant
LMC objects, namely,  \ion{H}{2} regions and Supernova remnants:
column (1) lists the atom; in columns (2) and (3) we give the average
elemental abundances for the symmetric and asymmetric PNe of our samples, 
in the usual form 12+log (N/H), with the sample size in parenthesis;
column (4) gives the LMC \ion{H}{2} region average from \citet{lei96},
as used in our Figures; columns (5) and (6) give the 
abundance ranges found by \citet{rus90} respectively in LMC \ion{H}{2} regions
and LMC Supernova remnants. The ranges are only indicative of the abundance distribution on the LMC, since they include a few observed objects for 
each diagnostics.

The ratio of the average neon abundance for asymmetric to 
symmetric PNe in the LMC, 1.7, is similar to that found in 
the Galaxy, 1.6 \citep{cor95}. Note, however, that the Galactic 
average is based on a sample that is markedly biased toward the symmetric 
PNe (see $\S$1). A similar overabundance in asymmetric PNe is found for 
sulfur and argon, although in these cases this 
includes a few PNe that do not follow the abundance-morphology relation.

Figure 3 shows the oxygen distribution among different Population PNe. 
Oxygen yield should be invariant with respect to progenitor mass or other 
parameters, at least in the LMC \citep{van97}. Data shown in Figure 3 
are consistent with the expectation that the oxygen abundance 
are within the same range in symmetric and asymmetric PNe. 
These results from our limited sample suggest that progenitors of 
asymmetric PNe were formed in an enriched environment that is typical of a 
very young stellar Population, while precursors of symmetric PNe formed in 
a medium that was under-abundant with respect to the LMC \ion{H}{2} region
average. 
PNe split into two distinct Population groups 
according to alpha-element (neon, argon, sulphur) abundance and morphology. 

In order to compare our findings with the analysis of Galactic PNe,
we looked for spatial segregation of symmetric and asymmetric PNe.
Although the sample size is modest, we did not find such a strong spatial 
segregation of LMC PNe across morphological type, or of their 
location with respect to the LMC bar or any other particular region, with the 
exception of a very slight predominance of asymmetric PNe along, rather than 
perpendicular to, the bar. This result is consistent with the short dynamical
mixing time of the LMC, and we should not expect a marked segregation of 
stellar Populations 
as is found, for example, in spiral galaxies. 

A number of theoretical models \citep{ren81,ibe83,van97} predict that 
lower-mass PN precursors typically go through the carbon stars phase, while 
the hot-bottom burning (HBB) takes place in higher mass precursors and
prevents carbon star formation. This aspect of stellar evolution does not
depend dramatically on the initial composition, and in particular for LMC 
PNe one expects the same carbon depletion and nitrogen enrichment above 
$\approx$4 \msun\ of main sequence stars \citep{van97}. Observations of Galactic PNe support 
the theory, in that there seem to be two AGB star types characterized by 
distinct carbon-enriched or depleted dust \citep{tra99}.

Then how do carbon and nitrogen abundances correlate with morphology in the
LMC? 
We plot nitrogen against carbon abundance 
in Figure 4. It is clear that symmetric 
PNe are well confined in such a plot, and all 
symmetric PNe are carbon-enriched 
with respect to the LMC \ion{H}{2} region average. 
The situation for asymmetric PNe is 
rather different: they are al nitrogen-enriched, yet three of them 
are also carbon-enriched with respect to the \ion{H}{2} region average. 
The figure could be interpreted as follows: low mass stars ($<4$ \msun\ on 
main sequence) go through the carbon star phase, and do not produce asymmetric 
PNe. The high mass stars do not go through the carbon star phase, they 
suffer HBB on the AGB, and they produce asymmetric PNe. Some of the 
low-mass stars producing carbon stars also end up as asymmetric PNe, perhaps 
through the common envelope phase. 

A discussion of the formation mechanisms for asymmetric PNe is in order, 
in light of our findings. 
If asymmetry in PNe were due uniquely to common envelope evolution, or
binary evolution in general, we would not expect to find any of the 
separations among morphological classes that we show in Figures 1 though 
3. That is, we do not expect the incidence of close binaries to vary 
as a function of the mass of the PN progenitor. 
Though this is a qualitative statement, it is substantiated by 
other observations (i.~e., the existence of bipolar PNe that do not show 
an equatorial ring, such as the Galactic PN Hubble 5), 
and studies of the initial mass 
function. 
On the other hand, there may be a small fraction of asymmetric PNe that are
developed as a consequence of close binary evolution, and this is also 
supported by observations. 
It may be that we are very far from understanding how the morphology of 
PNe is created, as there are a large number of variables in this game. 
What we can conclude with some certainty is that asymmetry in PNe is related 
to the Population type, and by inference the mass of the progenitor star. 
Any model for the formation of PNe that predicts the morphology must be 
consistent with this relationship. 

\section {Abundance Uncertainties and Systematic Effects}

If the results of the previous sections are borne out by additional 
observations and analysis, the effect on subsequent interpretations of 
PN formation and evolution could be substantial. 
In this section we take a critical look at the derivation of the chemical 
abundances that are the underpinning of the conclusions presented here. 

The first step is to give some information on the abundance uncertainties
as derived from the original papers. None of the references that we have used 
for abundance cite individual errors. \citet{dop97} do not discuss errors, 
but from the errorbars in 
their Fig. 7 it is possible to infer that the N, O, and C abundances
are good to 0.1 dex, while the errors on the alpha-elements are about 
0.08 dex. We should infer similar uncertainties for \citet{dop91a,dop91b},
since the three papers use the same abundance determination method.
\citet{mon88} evaluate that their oxygen and neon abundances are good to 0.1 or
0.15 dex, while neon and argon abundances are more uncertain, up to 0.2 dex.
Finally, \citet{lei96} determine that most elemental abundances are good
to 0.1 dex, with the exception of nitrogen, whose uncertainty is up to 0.2
dex or more. 

The potential for systematic errors from using different sources of 
abundance in the literature is worrisome enough to warrant some scrutiny. 
We compared the abundances from different bibliographic sources, and found 
that the values from \citet{dop91a,dop91b} and \citet{mon88} agree, within the 
quoted errors, with the abundances by \citet{lei96}. Plotting
the results of pairs of references shows only a scatter in the final results 
for most elements with no systematic differences.
The only element that is worrisome 
is nitrogen. \citet{lei96} give a nitrogen abundance that is 
systematically about 0.3 dex higher than other authors. 
However, omitting the work of any one paper does not change the
qualitative results of Fig. 4. In particular, any discrepancies do not
correlate with the ionization states of the nebulae\footnote{Low
ionization PNe are those in which O$^+$ is the dominant ionization state, while
in high
ionization PNe the O$^{++}$ state dominates.}. 

Most abundances are derived from observations using some form of the ionization
correction method to convert from measured ionic abundances to total
abundances. An ionization correction ``factor'' (ICF) for unseen 
ionization states is
required in this conversion.
\citet{ale97} found that the ICFs are very large in the case 
of low ionization nebulae, giving, for example, artificially high neon or
sulphur abundances
that in principle could alter the effects seen in Figures 1 though 3. 
We explored the original line intensity lines used for abundance calculation
by
\citet{lei96}, and \citet{dop91a,dop91b}, and we related the ionization level from 
the [\ion{N}{2}] to \ha line ratio, 
to the morphological type. The question becomes whether those classes of
PNe with segregated abundances have systematically low or high ionization.
We found no correlation between ionization level and morphology in our 
combined sample. Most of the nebulae show high ionization, and the few 
low-ionization objects are equally distributed among symmetric and 
asymmetric PNe. The only exception to this rule concerns extremely 
bipolar PNe (indicated with a filled square in the Figures). Among four 
of such PNe, three are low ionization. Eliminating these extreme bipolars 
from the plots would not change the conclusions of this paper. 
It is worth mentioning that \citet{ale97} compared ICF abundances to those
of model computation, and they find that Ne/H is the most reliable measure 
of the abundance of primordial elements since the O++ and Ne++ volumes are
very similar and the corrections for unseen ionization states are 
relatively small.

In general, LMC PNe are point-sources when observed from the ground, 
and the line 
intensities quoted in the literature typically refer to the 
global volume of the nebulae, thus the ICF problem here is minimal
\citep{ale97}. In fact, often Magellanic Cloud PNe are
discovered via [\ion{O}{3}] imaging (or [\ion{O}{3}] must have been present 
in an objective prism spectrum), so fewer PNe in our sample have low ionization.

One last concern comes from the possibility that the line intensities
used for abundances calculation
suffer from alteration due to the presence of a shock front. This is 
particularly worrisome for the sulphur abundances. In fact, artificially high
sulphur abundances may derive from excessively high intensities of the
low-ionization states of sulphur. 
We have checked the line
intensities for all PNe with 12+log(S/H)$>$7 to obtain 
diagnostic ratios for shocks, as 
explained by \citet{vei87}. 
In our sample, only two PNe have log~[\ion{O}{1}]($\lambda$6300)/\ha and 
log~[\ion{S}{2}]($\lambda$6716+6731)/\ha close to the limit for shock fronts
(see the dashed-dotted lines in \citet{vei87}'s Figs. 4 
through 6). We conclude that only a negligible fraction of the
asymmetric PNe in Figure 1 may have high sulphur abundance due to 
the presence of a shock front. The overall scientific results that we
describe in the previous chapter holds even if this was the case.

In conclusion, we feel that the results shown in this paper are quite sound, 
in spite of the inhomogeneous abundances. 
This work will be extended to the SMC in which the metal abundances
are considerably lower than in the LMC. In addition, we are extending the
ground-based spectroscopy to a far larger sample of LMC and SMC PNe.
Both the numbers of targets and the accuracy of the data will be greatly
improved as a consequence. 

\acknowledgements 

Thanks to Max Mutchler for his work on our LMC images, to Karen
Kwitter for carefully reading the manuscript, and to an anonymous referee
for important comments.
Support for this project was provided by NASA through grant number GO-08271.01-97A
from Space Telescope Science Institute, which is operated by the 
Association of Universities for Research in Astronomy, Incorporated, 
under NASA contract NAS5--26555.

\clearpage

\begin{table}
\caption{Average elemental abundances in the LMC}
\begin{tabular}{lccccr}
\hline
Atom& Symm.~PNe& Asymm.~PNe& \ion{H}{2}~Reg.&  \ion{H}{2}~Reg. & SNR\\
\hline
\hline\
\noindent

He& 	11.0 (19)&	11.0 (18)&	10.9&	10.91-11.03& n.a.\\
C&	8.76 (11)&	8.23 (7)&	7.87&	n.a.&	7.66\\
N&	7.83 (19)&	8.26 (18)&	6.97&	6.85-7.27& 7.26-7.45\\
O&	8.30 (20)&	8.41 (18)&	8.38&	8.18-8.60& 8.10-8.54\\
Ne&	7.49 (19)&	7.73 (18)&	7.64&	7.56-7.78& 7.11-7.95\\
S&	7.04 (15)&	7.15 (13)&	6.67&	6.68-7.0& 6.4-7.0\\
Ar&	6.08 (17)&	6.32 (15)&	6.20&	5.8-6.37& 6.51-6.65\\

\hline\
\end{tabular}
\tablecomments{Abundances are given as 12+log(N/X).
Numbers 
in parenthesis in columns (2) and (3) indicate the available 
PN sample for a given statistics.
\ion{H}{2} region abundances in column (4) 
are from \citet{lei96}. Abundance ranges for \ion{H}{2} Regions and Supernova
Remnants in columns (5) and (6) are from \citet{rus90}.}
\end{table}

\clearpage

\figcaption{S vs. Ne abundance of LMC PNe for morphological types Round 
({\it open circles}), elliptical ({\it asterisks}), quadrupolar 
({\it filled triangles}), bipolar core ({\it filled circles}), and 
bipolar ({\it filled squares}). The large crossed circle represents
the average for LMC \ion{H}{2} regions (see Table 1 for abundance ranges 
in \ion{H}{2} regions and SNR in the LMC).}

\figcaption{Argon vs. neon abundance, symbols as in Fig. 1.}

\figcaption{Oxygen vs. neon abundance, symbols as in Fig. 1.}

\figcaption{Nitrogen vs. carbon abundance, symbols as in Fig. 1.}

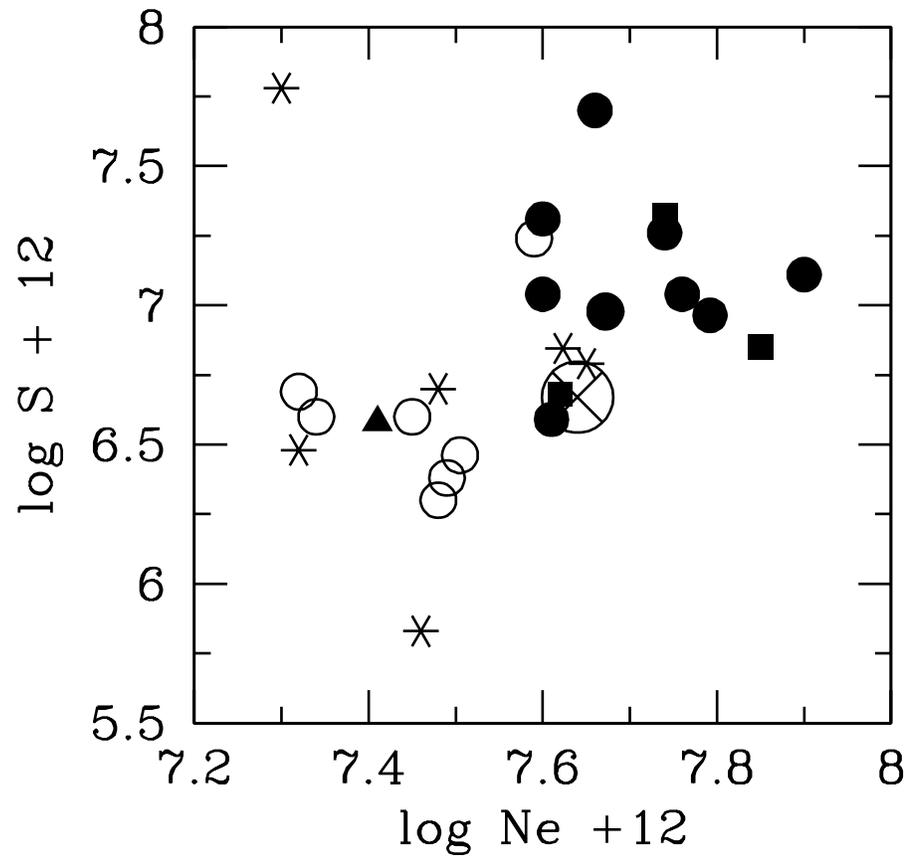
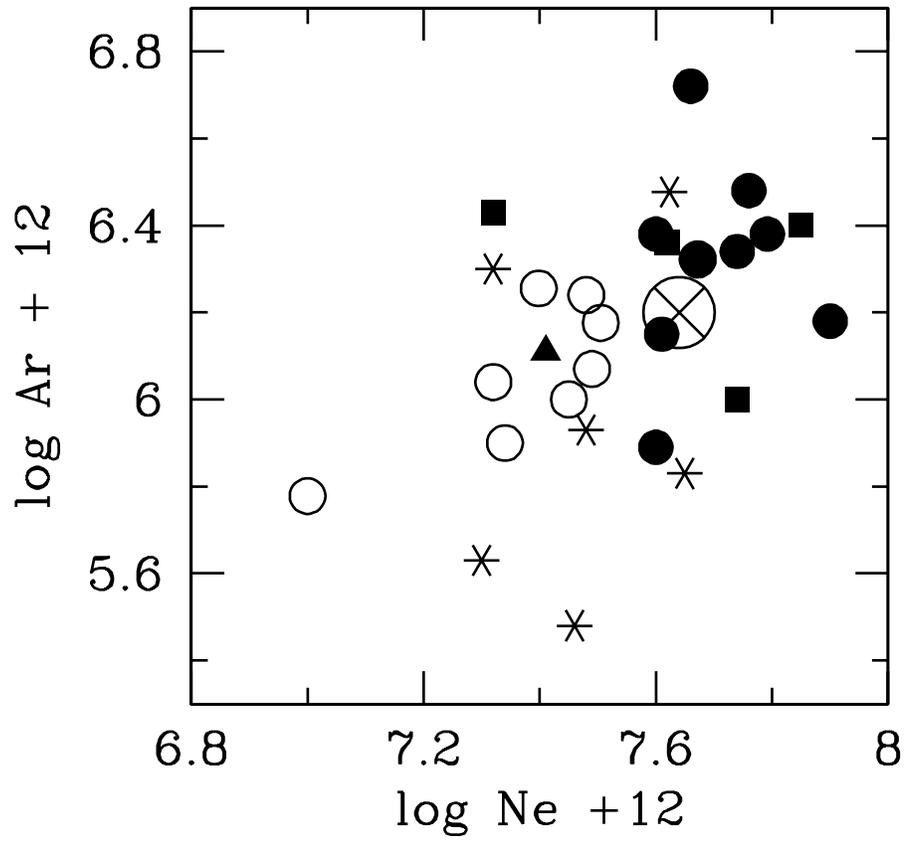
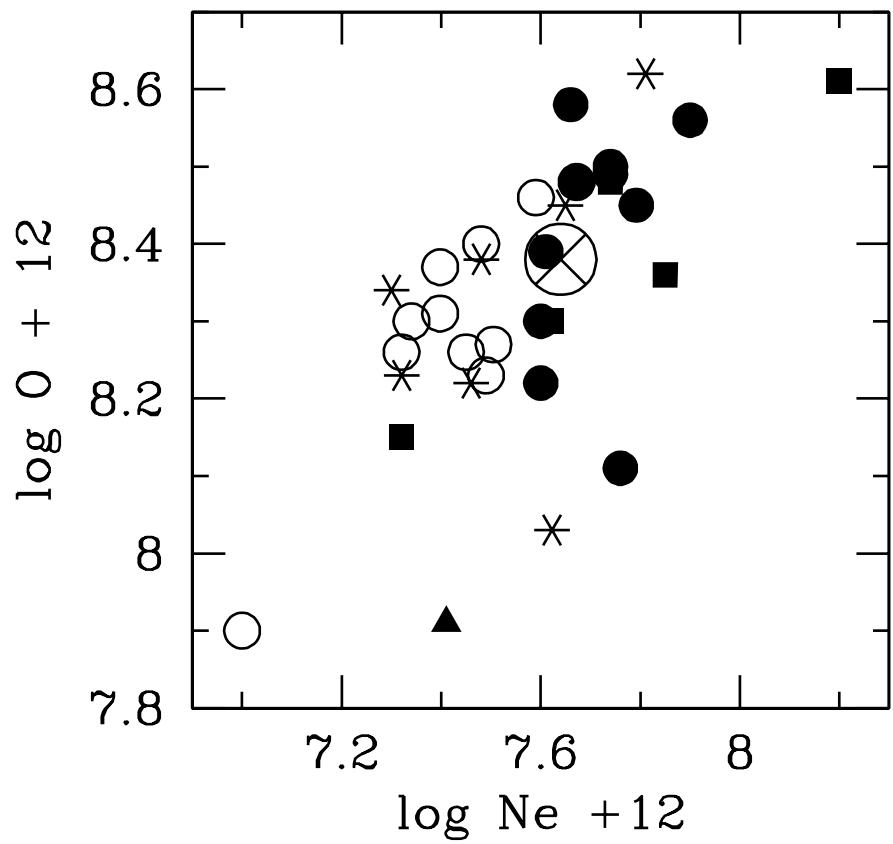
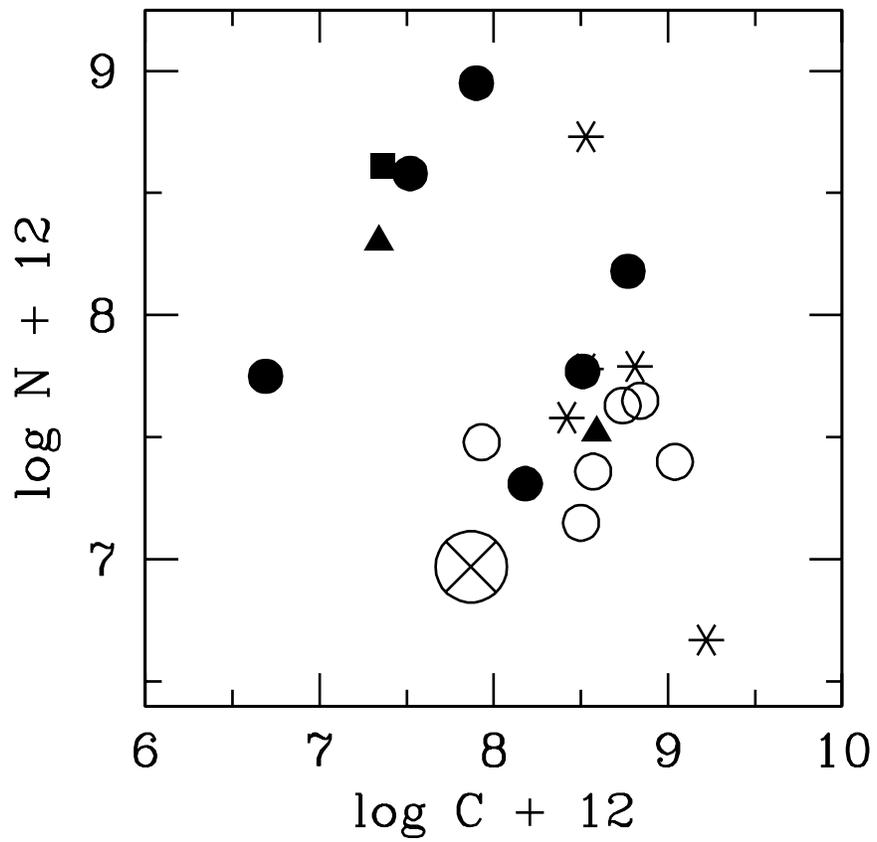
\end{document}